# Statistical properties and error threshold of quasispecies on single-peak Gaussian-distributed fitness landscapes

Duo-Fang Li [a], Tian-Guang Cao [a], Jin-Peng Geng [a], Jian-Zhong Gu [b], Hai-Long An[a], Yong Zhan [a*]

[a] *Institute of Biophysics, School of Sciences, Hebei University of Technology, Tianjin 300401, China*
[b] *China Institute of Atomic Energy, P. O. Box 275 (10), Beijing 102413, China*

**Abstract**: The stochastic Eigen model proposed by Feng et al. (Journal of theoretical biology, 246 (2007) 28) showed that error threshold is no longer a phase transition point but a crossover region whose width depends on the strength of the random fluctuation in an environment. The underlying cause of this phenomenon has not yet been well examined. In this article, we adopt a single peak Gaussian distributed fitness landscape instead of a constant one to investigate and analyze the change of the error threshold and the statistical property of the quasi-species population. We find a roughly linear relation between the width of the error threshold and the fitness fluctuation strength. For a given quasi-species, the fluctuation of the relative concentration has a minimum with a normal distribution of the relative concentration at the maximum of the averaged relative concentration, it has however a largest value with a bimodal distribution of the relative concentration near the error threshold. The above results deepen our understanding of the quasispecies and error threshold and are heuristic for exploring practicable antiviral strategies.

## 1. Introduction

The researches on the species evolution have been a focus for a long time. In the early 1970s, Eigen and Crow et al., based on Darwin's principle of nature selection, established respectively the Eigen model and the Crow-Kimura model to describe the evolution of the macromolecules (Eigen, 1971; Crow and Kimura, 1970). The Eigen model is a coupled mutation-selection model which describes the evolution of error-prone replicating molecules (Eigen, 1971). The Crow-Kimura model is known as a parallel mutation-selection model in which selection and mutation are two independent processes (Saakian et al., 2008). The Eigen model and Crow-Kimura model have appeared to be useful in understanding the origin and evolution of life. Their dynamics and equilibrium properties have been extensively studied in the past 40 years (Eigen et al., 1988 and 1989; Saakian et al., 2004; Tarazona, 1992; Nilsson and Snoad, 2000). The Eigen model has two important predictions: the quasispecies (Eigen and Schuster, 1977 and 1978) and error threshold (Eigen, 1971). The former means the distribution of some mutant sequences centered on the master sequence in equilibrium. The latter is a critical mutation rate above which all the macromolecule sequences lose their genetic information and are randomly distributed in the sequence space. The quasispecies and error threshold have been also confirmed in the evolution experiments of some viruses (Fishman and Branch, 2009; Wain, 1992; Domingo et al., 1992; Steinhauer et al., 1989).

In fact, the classic Eigen model was initially developed on the condition of infinite asexual population. To be closer to the reality, the theory was extended to the case of finite populations and diploid organism (Nowak and Schuster, 1989; Saakian and Hu, 2006; Wiehe et al. 1995). More realistic models were established by considering real biological effects, including genome recombination and repair effect, host immune response, complementation during coinfection and so on (Boerlijst et al., 1996; Tannenbaum and Shakhnovich, 2004; Kamp, 2003; Sardanyes and Elena, 2010). Furthermore, an empirical fitness landscape was introduced into the evolution model, which provided a broad framework of evolutionary dynamics for microbial population and cancer cells (Sardanyes et al., 2014; Rozen et al., 2002). Recently, the randomized study on biological processes both in theory and experiment has been quite active (Guo and Mei, 2014; Domingo et al., 1978). The stochastic versions of the Eigen model were built by physical methods, such as the Langevin equation and Markov process. The stochastic dynamics of species evolution were elucidated in these random models (Galstyan and Saakian, 2012; Inagaki, 1982). Nevertheless, the stochastic characteristics on the Eigen model have not been well understood. The species evolution for a real population, especially at the molecular level, is inevitably affected by various random factors, such as genetic mutations and environmental fluctuations (Neher and Shraiman, 2012). Therefore, the important physical parameters appearing in species evolution models are also subject to various random factors and should be stochastic. The foot-and-mouth disease virus (FMDV) evolution experiments found that the relative fitness of the virus appears a fluctuating pattern around a constant average fitness (Lazaro and Escarmis, 2003).

Recently, the random Eigen models in which the deterministic physical parameters were treated as Gaussian distributed random variables were proposed (Feng et al., 2008; Qiao et al., 2014). But only qualitative conclusions were given in these randomization studies. The reason for the change of the error threshold has been not yet clear. The randomization effects of the Eigen model are still worth studying. Firstly, how to measure the extension of the error threshold and what is the relationship between the extension and the fluctuation strength? Secondly, when a deterministic concentration is replaced by an ensemble of concentrations, how does the ensemble pass through the error threshold?   The above questions actually motivate the present work.

In this article, the change of the error threshold and the statistical properties of the relative concentrations of the quasispecies are studied on a single peak Gaussian distributed fitness landscape. In section 2, the classic Eigen model and the random Eigen model are introduced, respectively. In section 3, the Gaussian distributions of random fitness of the mutant sequences and master sequence are firstly tested, then a quantitative analysis for the extension of the error threshold is performed, and finally the distribution characteristics of the quasispecies, especially those around the error threshold are examined based on the random Eigen model. In section 4, the main conclusions of this work are drawn.

## 2. Models

*The Eigen model*    In the Eigen model, the individuals are specified by the sequences of $N$ digits of basis κ. If one only considers purine and pyrimidine, $\kappa = 2$. The total number of possible sequences is $2^N$. The difference between arbitrary two sequences is represented by the Hamming distance. Those sequences with the same Hamming distance from the master sequence are combined into a class. As a result, there are $N+1$ classes of the sequences, $I_0, I_1, \cdots, I_N$. $I_0$ is the master class, $I_i (i>0)$ means a mutant class. The relative concentration of each class in the population, $x_i$, satisfies the following equation:

$$\frac{dx_i}{dt} = \sum_{j=0}^{N} f_j q_{ij} x_j - \phi(\vec{x}) x_i .\qquad(1)$$

Here, $f_i$ is the fitness representing replication rate of class $I_i$ and $q_{ij}$ is the transition probability from class $I_j$ to class $I_i$. $\phi$ is dilution flux, which satisfies the condition of $\phi(\vec{x}) = \sum f_i x_i$. Eq. (1) can be converted from the nonlinear differential form into a linear differential one by a variable transformation (Thompson and McBride, 1974). The equilibrium distribution of the population could be obtained by the eigenvalue of the coefficient matrix W ($W_{ij} = f_j \cdot q_{ij}$) (Jones et al., 1976). The largest eigenvalue and its corresponding right eigenvector of the matrix W respectively give the production rate and the absolute concentration of each class ($X_i$) in the equilibrium state. The normalized eigenvector, $x_i = X_i/X$ represents the relative concentration of each class in population and X is the sum of all $X_i$.

In the deterministic Eigen model, the fitness function commonly adopted a single peak fitness landscape which assumes all sequences in a given class have exactly the same properties. Its mathematical form is written as:

$$f_0 = A_0, \quad f_i = A_i = A_1 < A_0 \ (i > 0) \qquad(2)$$

Here, A0 and A1 are constants. And the uniform mutation rate is assumed, which means that only point mutation is considered and the mutation rates at different sites are the same and independent of each other (Swetina and Schuster, 1982; Nowak and Schuster, 1989). We may take

$$q_{ij} = \sum_{l=l_{min}}^{l_{max}} \binom{j}{l}\binom{N-j}{i-l} q^{N-j-i+2l}(1-q)^{j+i-2l} . \qquad(3)$$

Here $q$ represents copying fidelity of each site of a macromolecule sequence, and mutation rate is then $\mu = 1 - q$. $l_{min} = \max\{0, j+i-N\}$, and $l_{max} = \min\{j, i\}$. In Eq. (2) and Eq. (3), all parameters are deterministic.

*The random Eigen model* Based on the assumption of single peak fitness, the deterministic fitnesses are replaced by Gaussian distributed random variables in the present work, and the probability density distributions of the random fitnesses $y_i$ are given as follows:

$$P(y_i) = \frac{1}{\sqrt{2\pi\omega_i^2}} e^{\frac{-(y_i - \overline{y_i})^2}{2\omega_i^2}} \qquad i=0, 1, 2, 3 \dots N. \qquad(4)$$

Here $\overline{y_i}$ and $\omega_i^2$ denote the average values and variances of the random fitnesses. Thus, the fitness given by the deterministic Eigen model is replaced by the above single peak Gaussian distributed fitness landscape (spGDFL). To facilitate the comparison with the deterministic model, the average values of the master class fitness and mutant class fitnesses in the numerical simulations are taken to be $A_0$ and $A_1$ respectively. The fluctuation strength of each random variable is measured by $d_i = \omega_i/\overline{y_i}$. Without loss of generality, we take $d_i = d$. It is worth noting that there exists an upper limit for the fluctuation strength of the random variables ($d = 0.25$ in the present work) in numerical simulation. Beyond this limit, the system is unstable. Generally speaking, the fitness fluctuation in practice is not so large.

## 3. Results

Throughout the calculation, the length of the macromolecules sequence is N=20. The single peak fitness landscape is set to $A_0 = 10$, $A_1 = 1$. The number of realizations of random sampling is 10000. The fluctuation strength $d$ is taken to be 0.05, 0.10, 0.15, 0.20 and 0.25 respectively. To ensure the reliability of the random fitnesses, their probability density distributions are calculated. The results show that for different values of the fluctuation strength, the random fitnesses follow the Gaussian distributions, and the goodness of fitting for all of the random variables is above 0.99. By numerical simulations, the relative concentration ensemble for each class is generated. Based on the ensemble, the features of the error threshold and the quasispecies in the random Eigen model can be investigated.

### 3.1 Characteristics of the error threshold in the random Eigen model

For the deterministic Eigen model ($d = 0$), Swetina and Schuster (1982) obtained the distribution of the relative concentration of each class versus the mutation rate, as shown in Fig. 1(a). Number 0 in the figure represents the master class, and 1, 2… N denote different mutant classes. In the deterministic Eigen model with a single peak fitness landscape, the evolution process of the population with mutation rate is a mixing process between the master and mutant sequences. The master sequence melts gradually in the mutant sequences with increasing mutation rate and completely dissolves into the mutant sequences at the error threshold. The error threshold for *N*=20 is located at mutation rate $\mu = 0.112$, which is a sharp point similar to a phase transition in physics. Over the error threshold, the sequences have negligibly small concentrations (roughly speaking, zero concentrations) and the complementary classes get together because of the same degeneracy. Here complementary classes are the two classes $I_i$ and $I_j$ with $i + j = N$.

In the random Eigen model, the averaged relative concentrations for each class ensemble versus increasing mutation rate are computed and shown in Fig. 1(b) (*d*=0.1) and Fig. 1(c) (*d*=0.2). The error threshold in the random Eigen model becomes a smooth crossover region. For instance, the crossover region in the case of $d = 0.1$ is located between 0.112 and 0.119. The crossover region becomes wider as the fluctuation strength increases. One can understand the above phenomenon as follows. The randomization of the fitness landscape modifies the relative concentration of a sequence from a specific value into an ensemble consisting of various relative concentration values for a given mutation rate value. The probability distribution of those concentration values appears as a concentration wave packet with certain width. The concentration wave packets with certain widths facilitate their mixing, which results in the occurrence of the dissolution at a lower mutation rate than the error threshold. This implies that the error threshold extends downward. On the other hand, above the error threshold, the concentration wave packets with certain widths have some components with non-zero concentrations. This implies that the error threshold extends upward. Therefore, the extension of the error threshold completely results from the fitness landscape randomization. With the increment of the fluctuation strength of the fitness landscape, the wave packets get wider and wider, and the range of the crossover region becomes larger and larger. That is to say the width of the error threshold increases with the fluctuation strength.

The upper limit of the crossover region obviously surpasses the error threshold given by the deterministic Eigen model. This fact should be considered when dealing with the practical problems of species evolution. Although the error threshold changes significantly in the random Eigen model, the relative concentrations in the other regions are basically consistent with those in the deterministic Eigen model, implying that they are relatively stable against the fitness fluctuation.

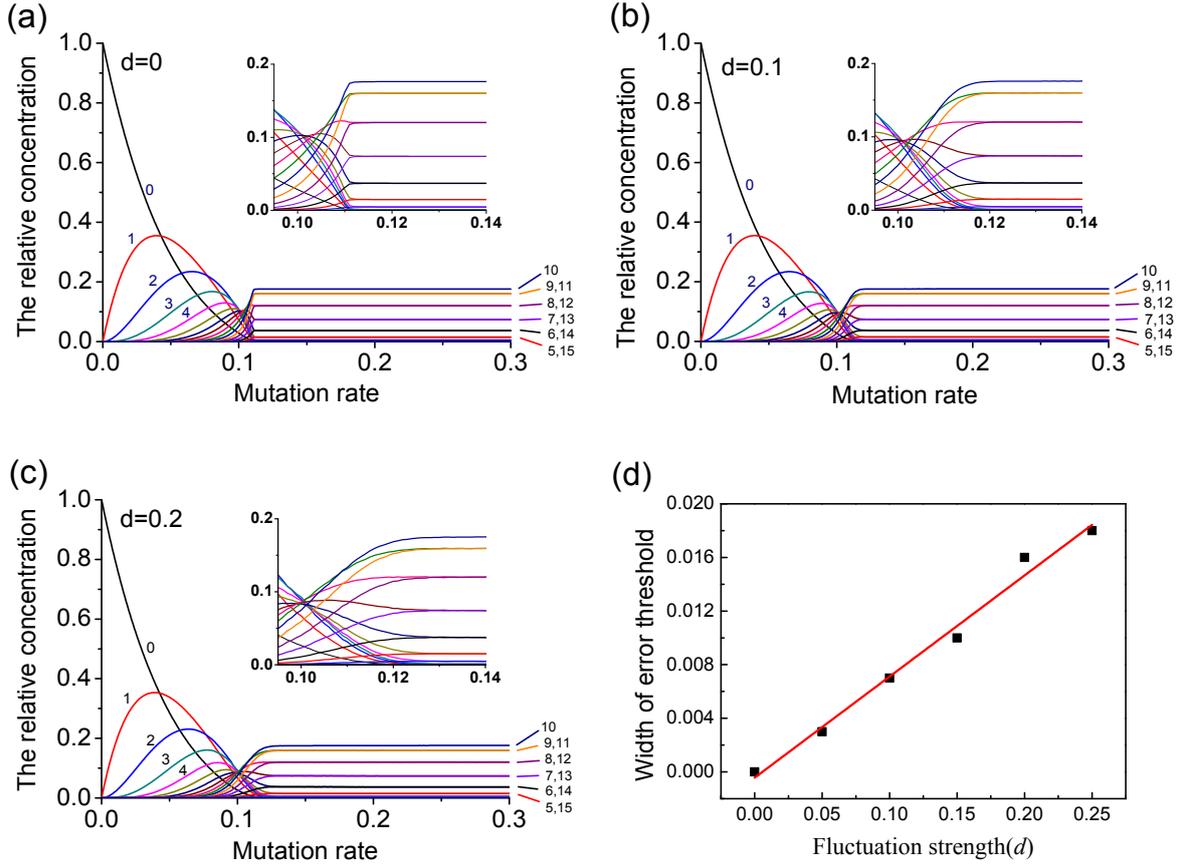

Figure1. The relative concentrations of each class versus the mutation rate in equilibrium. **(a)** The distribution of relative concentration in the deterministic Eigen model ($d = 0$). **(b)** and **(c)** Those in the random Eigen model with $d = 0.1$ and $d = 0.2$ respectively. **(d)** The relationship between the width of the crossover region and the fluctuation strength of the random variables is basically linear.

To measure quantitatively the broadening effect of the error threshold, the range of the crossover region is defined. The starting point of the crossover region is the error threshold given by the deterministic Eigen model. And the end point is the position where the relative difference in the relative concentration of two complementary classes is less than 0.01. The relative difference $c$ is given by

$$c = \frac{x_i - x_j}{(x_i + x_j)/2} \qquad (5)$$

For different complementary classes, their relative concentrations behave in a similar manner in the crossover region. According to the above definition of the crossover region, the width of the crossover region versus the fitness fluctuation strength is determined and shown in Fig. 1(d). It is shown that the relation between the width of the crossover region and the fitness fluctuation strength is approximately linear. We notice that when d changes from 0 to 0.25 the maximum value of the error threshold width is about 0.018 and much smaller than the error threshold 0.112 (their ratio is 16%). The fluctuation strength indicates a kind of perturbation, and the error threshold width measures the response of the system to the perturbation. It should be noted that the linearity is obtained in the conditions: (1) the Eigen model with a randomized single-peak fitness landscape; (2)

0<d<0.25, which sets the limits of applicability of the linearity. The result has an important implication for antiviral strategies. In order to completely drive viruses extinct, the mutation rate must be enhanced up to the upper limit of the crossover region.

**3.2 The statistical fluctuation and distribution of the relative concentration for the given quasispecies**

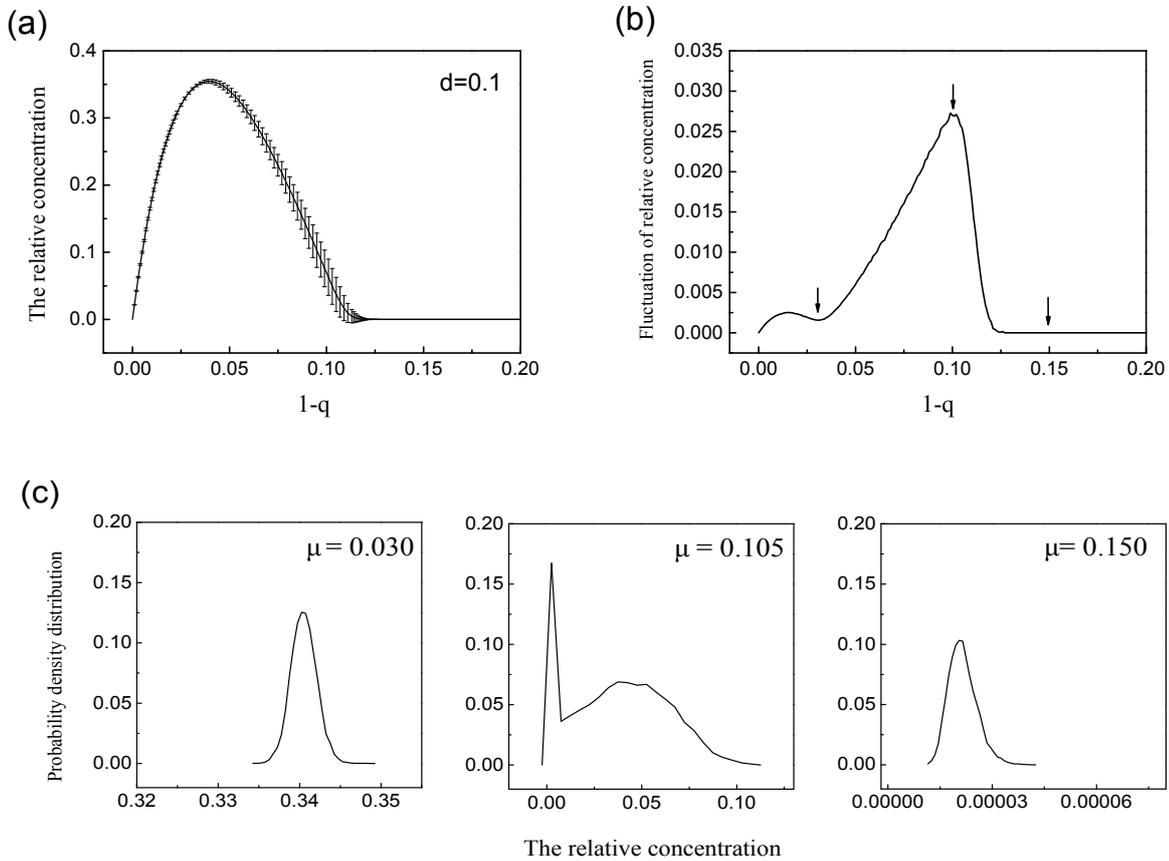

Figure 2. The statistical characteristics of the relative concentration of $I_1$ in random Eigen model with $d = 0.1$. (**a**). The averaged value and standard error of the relative concentration of class $I_1$. (**b**). The concentration fluctuation of class $I_1$ versus mutation rate. (**c**). The probability density distributions of the relative concentration of class $I_1$ with values of mutation rate $\mu = 0.030$, 0.105 and 0.150.

The fitness randomization turns the relative concentration of each class from specific values into various ensembles. The ensemble averaged relative concentrations are related to actual observed concentration. In viral evolution experiments, the relation between the statistical properties of a mutant class and the sequence population evolution is of widespread concern. The statistical properties of the relative concentrations of the classes have been investigated in the present work. We here take mutant class $I_1$ as an example to illustrate the situation. The ensemble averaged values and standard error of class $I_1$ with $d = 0.1$ are displayed in Fig. 2(a). To more clearly display the fluctuation for different values of mutation rate, in Fig. 2(b), we show how the standard error of $I_1$ changes as mutation rate increases. It can be seen that the fluctuation takes its minimum value when the averages relative concentration of class $I_1$ has its maximum value, and it moves into maximum value as mutation rate gets close to the error threshold. The standard error of class $I_1$ nearly vanishes when mutation rate goes beyond the error threshold. It is therefore suggested that the classes are relatively stable when mutation rate is below the error threshold, and they are unstable in the vicinity of the error threshold in the case of the spGDFL.

The probability density distributions of the relative concentration of class $I_1$ are given in Fig. 2(c) for three different values of mutation rate, which correspond to the minimum and maximum of the concentration fluctuation and the situation beyond the error threshold. The distributions at minimum of concentration fluctuation and beyond the error threshold are basically normal distributions. It is interesting that there is a bimodal distribution as mutation rate is close to the error threshold. The bimodal distribution may result from the large relative concentration fluctuation near the error threshold.

**3.3 The characteristics of bimodal distribution near the error threshold**

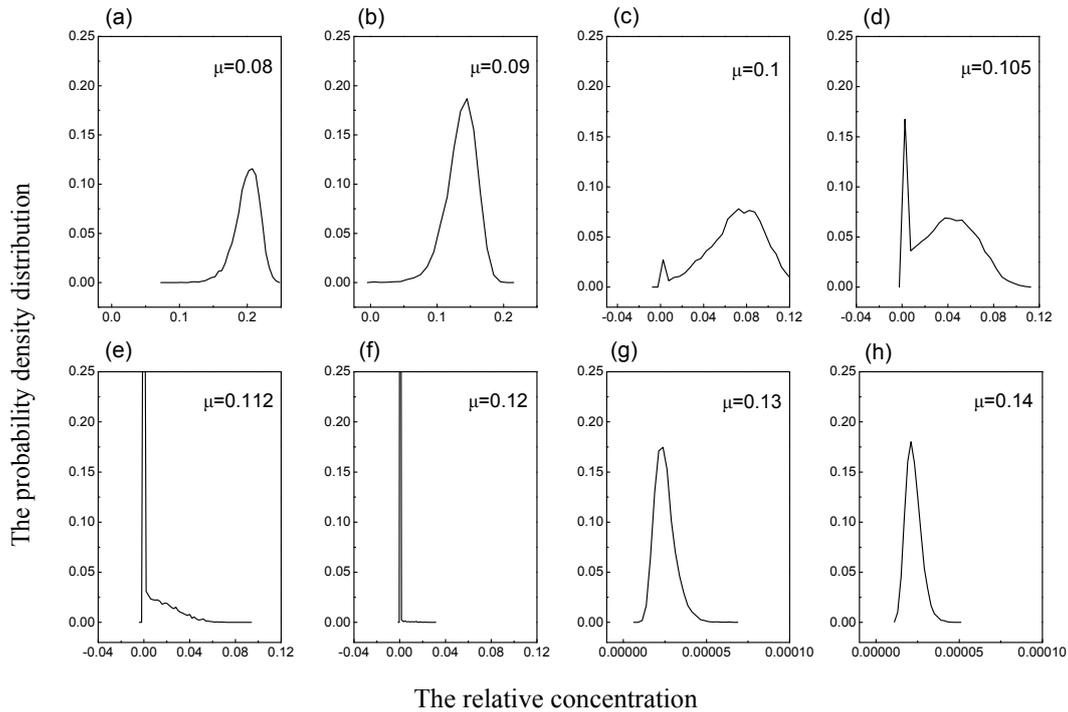

Figure 3. The detailed relative concentration distributions of class $I_1$ are given in the vicinity of error threshold at $d = 0.1$.

The distribution features of the relative concentration of class $I_1$ are further studied with a finer mutation rate interval in the vicinity of the error threshold which covers the crossover region ($\mu = 0.112 - 0.119$) as shown in Fig. 3 for the case of $d = 0.1$. It is seen that the bimodal distribution only survives in the region of 0.112-0.119 and turns into a single-peaked distribution when mutation rate goes beyond the error threshold region. Furthermore, the right peak of the bimodal distribution moves toward the left peak as mutation rate increases, and the left one has a fixed position and its peak value is growing with increasing mutation rate. The bimodal distribution implies the coexistence of the life and death. The right peak presents the life and the left one indicates the death.

In the deterministic Eigen model, the dynamical properties of the population are fully determined by the coefficient matrix **W** (Nowak, 2006). Therefore, the formation of the bimodal distribution may also be relevant with the properties of the matrix **W** even in the presence of the fitness randomization. In the deterministic Eigen model, the largest eigenvalue at the error threshold is a single value, denoted by $T_C$. Its corresponding right eigenvector represents the concentration distribution of the population at the error threshold. In the random Eigen model, the fitness randomization turns the largest eigenvalue of matrix **W** into various random values, which form an ensemble. The eigenvalue ensemble has some kind of distribution $T_G$, which is illustrated in Fig. 4. As mutation rate is close to the error threshold, the peak position of the distribution $T_G$ is approaching to $T_C$. The front part of the $T_G$ may reaches $T_C$ ahead. For a specific class, say class $I_1$, its partial concentration values become a negligible constant since the front part reaches or goes over $T_C$. When $T_G$ gradually moves toward $T_C$, more and more partial concentration values pass through the error threshold. As a result, the statistics counts for the partial concentration values with a negligible constant grow, and the left peak of the bimodal distribution becomes higher and higher. That is the reason for the formation of the bimodal distribution.

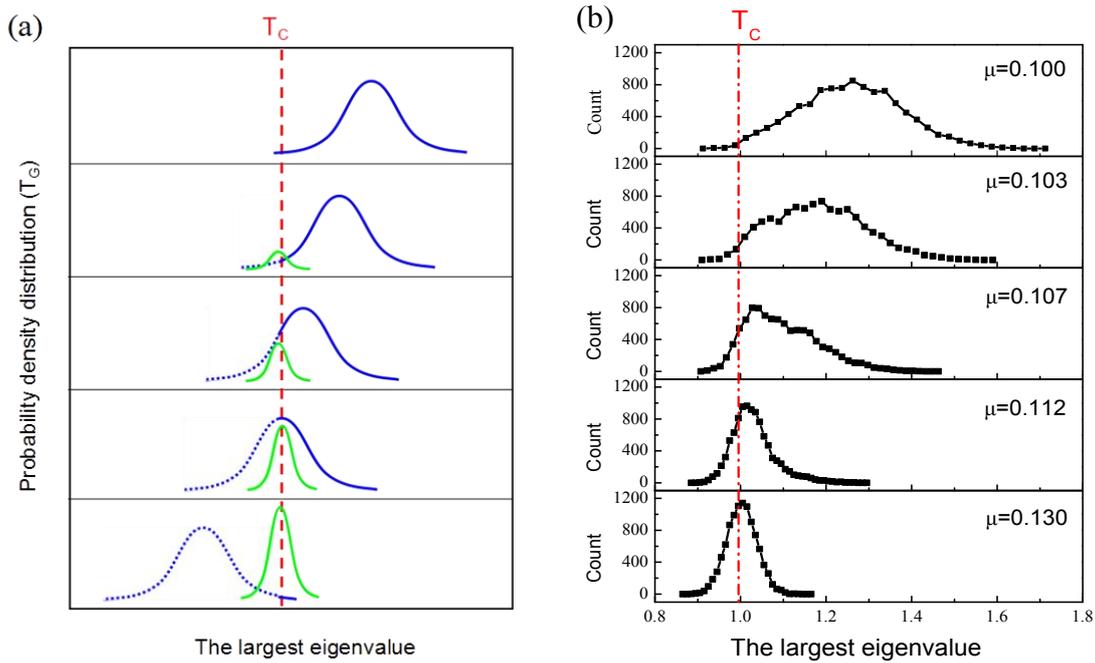

Figure 4. The schematic diagram and numerical (real) results of the largest eigenvalue ensemble $T_G$ around the error threshold. $T_C$ denotes the largest eigenvalue of the coefficient matrix **W** in the deterministic Eigen model at the error threshold. **(a)** The distributions of the largest eigenvalue ensemble (blue lines) passing through the error threshold, the passed parts (blue dotted lines) obey new distributions (green lines). **(b)** The real distribution of $T_G$ obtained by numerical simulations versus increasing mutation rate.

## 4. Discussions and conclusions

Nowadays, the Eigen model has gained the further development in theory and application. The emergence of the quasispecies proposes two different pathways to virus extinction: lethal mutagenesis and crossing the error threshold (Tejero et al., 2010). Based on this understanding, new antiviral strategies have been suggested for antiviral and cancer therapy (Eigen, 2002; Manrubia et al., 2010). The related viral experiments have been performed on the HCV (hepatitis C virus), HIV (human immunodeficiency virus), LCMV (lymphocytic choriomeningitis virus) and others (Domingo and Gomez, 2007). Although some useful results have been obtained in the study of the Eigen model, there is still a long distance from its practical applications. Randomization of the Eigen model brings it closer to the practical situation.

In this paper, we have adopted a single-peaked Gaussian distributed fitness landscape instead of a constant one to investigate and analyze the change of the error threshold and the statistical property of the quasi-species population. We found a roughly linear relation between the width of the crossover region and the fitness fluctuation strength. The fitness fluctuation can be induced by kinds of external factors, for instance, temperature and pH value, its strength might be controlled by external factors. Therefore, one may control the width of the crossover region externally based on the linear relation, which is certainly important for designing antiviral strategies. We have noticed that the mutagens were used to increase the mutation rate of viruses and cancer cells to achieve the antiviral and anticancer purpose (Loeb and Mullins, 2000; Graci and Cameron, 2006; Ruiz-Jarabo, et al., 2003). Because the error threshold turns into a crossover region in reality, the upper limit of the crossover region should be considered in order to completely eliminate viruses and cancer cells in the above studies.

Meanwhile, for a given class, the fluctuation of the relative concentration and the behavior of the relative concentration around the crossover region have been examined in detail. The fluctuation of the relative concentration has a minimum with a normal distribution of the relative concentration at the maximum of the averaged relative concentration, it has however a largest value with a bimodal distribution of the relative concentration near the error threshold. There have been some studies to begin to pay attention to the distribution of a single class in virus evolution experiments (Zhu et al. 2009). Although such experiments are very difficult to implement currently, we hope our above finding will be verified by future experiments. We explained the bimodal distribution based on the coefficient matrix. The bimodal distribution itself implies that the coexistence of the life and death could occur around the crossover region, which is a kind of residual life. It was also found that the relative concentration of a class dies out gradually.

Finally, we would say that the physical parameter randomization method used in the present work has some advantages over other approaches, such as the Langevin equation and Markov process, being simple and clarity in physical picture, and applies to other models for species evolution.

## Acknowledgements

The authors are greatly indebted to Professor Yizhong Zhuo for his discussions and suggestions. We thank the Natural Science Foundation of Hebei Province of China Grant No. C2013202192.

## References

Boerlijst, M.C., Bonhoeffer, S., Nowak, M.A., 1996. Viral quasi-species and recombination. Proc. R. Soc. Lond. B 263**,** 1577-1584.

Crow, J.F., Kimura, M., 1970. An introduction to population genetics theory. Harper and Row, New York.

Domingo, E., Escarmis, C., Martinez, M.A., Martinez, S.E., Mateu, M.G., 1992. Foot-and-mouth disease virus populations are quasispecies. Curr. Top. Microbiol. 176, 33-47.

Domingo E, Sabo, D., Taniguchi, T., Weissmann, C., 1978. Nucleotide sequence heterogeneity of an RNA phage population. Cell 13 (4), 735-744.

Domingo E., Gomez, J., 2007. Quasispecies and its impact on viral hepatitis. Virus Res. 127,131-150.

Eigen, M., 1971. Selforganization of matter and the evolution of biological macromolecules. Naturwissenschaften.

58 (10), 465-523.

Eigen, M., 2002. Error catastrophe and antiviral strategy. Proc. Natl. Acad. Sci. U.S.A. 99, 13374-13376.

Eigen, M., Schuster, P., 1977. A principle of natural self-organization Part A: emergence of the hypercycle. Naturwissenschaften 64, 541-565.

Eigen, M., Schuster, P., 1978. A principle of natural self-organization Part B: the abstract hypercycle. Naturwissenschaften 65, 7-41.

Eigen, M., McCaskill, J., Schuster,P., 1988. Molecular Quasi-species. J. Phys. Chem. 92,6881-6891.

Eigen, M., McCaskill J., Schuster, P., 1989. The Molecular Quasi-Species. Adv. Chem. Phys. 75, 149-263.

Feng, X.L., Li, Y.X., Gu, J.Z., Zhuo, Y.Z., Yang, H.J., 2007. Error thresholds for quasispecies on single peak Gaussian-distributed fitness landscapes. J. Theor. Biol. 246 (1), 28-32.

Fishman, S.L., Branch, A.D., 2009. The quasispecies nature and biological implications of the hepatitis C virus. Infect. Genet. Evol. 9 (6), 1158-1167.

Galstyan, V., Saakian, D.B., 2012. Dynamics of the chemical master equation, a strip of chains of equations in d-dimensional space. Phys. Rev. E 86, 011125.

Graci, J.D., Cameron, C.E., 2006. Mechanisms of action of ribavirin against distinct viruses. Rev. Med. Virol. 16 (1), 37-48.

Guo, W., Mei, D.C., 2014. Stochastic resonance in a tumor–immune system subject to bounded noises and time delay. Physica A 416, 90-98.

Inagaki., H., 1982. Selection under random mutations in styochastic eigen model. Bull. Math. Biol. 44, 17-28.

Jones, B.L., Enns, R.H., Rangnekar, S.S., 1976. On the theory of selection of coupled macromolecular systems. Bull. Math. Biol. 38(1), 15-28.

Kamp, C., 2003. A quasispecies approach to viral evolution in the context of an adaptive immune system. Microbes infect. 5 (15), 1397-1405.

Lazaro, E., Escarmis, C., 2003. Resistance of virus to extinction on bottleneck passages: Study of a decaying and fluctuation pattern of fitness loss. Proc. Natl. Acad. Sci. U.S.A.100, 10830.

Loeb, L.A., Mullins, J.I., 2000. Lethal mutagenesis of HIV by mutagenic ribonucleoside analogs. AIDS Res. Hum. Retrov. 16 (1), 1-3.

Manrubia, S.C., Domingo, E., Lazaro, E., 2010. Pathways to extinction: beyond the error threshold . Philos. T. Roy. Soc. B 365, 1943-1952,

Neher, R.A., Shraiman, B.I., 2012. Fluctuations of fitness distributions and the rate of Muller's ratchet. Genetics 191 (4), 1283-1293.

Nowak, M.A., 2006. Evolutionary Dynamics: Exploring the Equations of life. Canada: the President and Fellows of Harvard College. 34-35.

Nowak, M., Schuster, P., 1989. Error threshold of replication in finite populations mutation frequencies and the onset of Muller's ratche. J. Theor. Biol. 137, 375-395.

Nilsson, M., Snoad, N., 2000. Error thresholds for quasispecies on dynamic fitness landscapes. Phys. Rev. E 84 (1), 191-194.

Qiao, L.H., Zhao, T.J., Gu, J.Z., Zhuo, Y.Z., 2014. Eigen model of randomness in species evolution. Acta Phys. Sin. 63 (10), 108701.

Rozen, D.E., de Visser, J.A., Gerrish, P.J., 2002. Fitness effects of fixed beneficial mutations in microbial populations. Curr. Biol. 12 (12), 1040-1045.

Ruiz-Jarabo, C., Ly M.C., Domingo, E., Torre J.C., 2003. Lethal mutagenesis of the prototypic arenavirus lymphocytic choriomeningitis virus (LCMV). Virology 308 (1), 37-47.

Saakian, D.B., Hu, C.K., Khachatryan, H., 2004. Solvable biological evolution models with general fitness functions and multiple mutations in parallel mutation-selection scheme, Phys. Rev. E **70**, 041908.

Saakian, D.B., Hu, C.K., 2006. Exact solution of the Eigen model with general fitness functions and degradation rates. Proc. Natl. Acad. Sci. U.S.A. 103 (13), 4935-4939.

Saakian, D.B., Rozanova, O., Akmetzhanov, A., 2008. Dynamics of the Eigen and the Crow-Kimura models for molecular evolution. Phys. Rev. E 78 (4), 041908.

Sardanyés, J., Elena, S.F., 2010. Error threshold in RNA quasispecies models with complementation. J. Theor. Biol. 265 (3), 278-286.

Sardanyes, J., Simo, C., Martinez, R., Sole, R.V., Elena, S.F., 2014. Variability in mutational fitness effects prevents full lethal transitions in large quasispecies populations. Sci. Rep. 4, 4625

Steinhauer, D.A., Torre, J.C., Meier, E., Holland, J.J., 1989. Extreme heterogeneity in populations of vesicular stomatitis virus. J. Virol. 63 (5), 2072-2080.

Swetina, J., Schuster P., 1982. Self-replication with errors a model for polynucleotide replication. Biophysical chemistry 16, 329-345.

Tannenbaum, E., Shakhnovich, E.I., 2004. Error and repair catastrophes: A two-dimensional phase diagram in the quasispecies model. Phys. Rev. E 69 (1), 011902.

Tarazona, P., 1992. Error thresholds for molecular quasispecies as phase transitions:from simple landscapes to spin-glass models. Phys. Rev. A 45, 6038-6039.

Tejero, H., Marin, A., Montero, F., 2010. Effect of lethality on the extinction and on the error threshold of quasispecies. J. Theor. Biol. 262,733-741.

Thompson, C.J., Mcbride, J.L., 1974. On Eigen's theory of selforganization of molecules and the evolution of biological macromolecules. Math. Biosci. 21, 127-142.

Wain, H.S., 1992. Human immunodeficiency virus type 1 quasispecies in vivo and ex vivo. Curr. Top Microbiol. Immunol. 176, 181-193.

Wiehe, T., Baake, E., Schuster, P., 1995. Error propagation in reproduction of diploid organisms. A case study on single peaked landscapes. J. Theor. Biol. 177 (1), 1-15.

Zhu, Y., Yongky, A., Yin, J., 2009. Growth of an RNA virus in single cells reveals a broad fitness distribution. Virology 385 (1), 39-46.